\documentclass[a4paper,11pt,oneside]{article}
\usepackage[english]{babel}
\usepackage[dvips]{graphicx}
\usepackage{amsthm}
\usepackage{exscale}
\usepackage{amsfonts}
\usepackage{amssymb}
\usepackage[intlimits,sumlimits]{amsmath}
\usepackage{amsmath,cases}
\usepackage[latin1]{inputenc}
\usepackage{mathrsfs}
\usepackage{fancyhdr}

\theoremstyle{definition}

\newtheorem{thm}{Theorem}[section]

\newcommand{\xb}{\mathbf{x}}

\newcommand{\vb}{\mathbf{v}}
\newcommand{\wb}{\mathbf{w}}
\newcommand{\ub}{\mathbf{u}}

\newcommand{\omb}{\boldsymbol{\omega}}
\newcommand{\ud}{\mathrm{d}}

\newcommand{\unifSo}{u_{S^2}(\ud \boldsymbol{\omega})}

\newcommand{\intethree}{\int_{\mathbb{R}^3}}
\newcommand{\lnorm}{\mid\mid \!}
\newcommand{\rnorm}{\! \mid\mid}
\newcommand{\tlnorm}{\mid\mid\mid \!}
\newcommand{\trnorm}{\! \mid\mid\mid}
\newcommand{\rnorma}{\! \mid\mid_{\ast}}
\newcommand{\rnormm}{\! \mid\mid_{_{1}}}

\newcommand{\rone}{\mathbb{R}}
\newcommand{\rthree}{\mathbb{R}^3}

\newcommand{\Lone}{\mathrm{L}^1(\mathbb{R}^3)}

\author{\textsc{By Emanuele Dolera} \\
\emph{Universit\`a di Modena e Reggio Emilia }}

\title{\textbf{Spatially homogeneous Maxwellian molecules in a neighborhood of the equilibrium}}

\date{}

\begin{document}
\maketitle
\begin{abstract}
{\footnotesize This note deals with the long-time behavior of the solution to the spatially homogeneous Boltzmann equation for Maxwellian molecules, when the initial datum belongs to a suitable neighborhood of the Maxwellian equilibrium. In particular, it contains a quantification of the rate of exponential convergence, obtained by simple arguments.}
\end{abstract}


\section{Introduction and main result}

This note deals with the convergence of the solution of the Boltzmann equation to the Maxwellian equilibrium, as time goes to infinity. More specifically, it aims at providing a quick, homogeneous and detailed proof that initial data belonging to a suitable neighborhood of the equilibrium produce solutions that remain in a slightly greater neighborhood, and converge exponentially fast. There is a large number of works about this kind of problems and the results contained in this note have been already formulated, more or less explicitly, in previous studies, where partial proofs have also been provided. On the one hand, a lot of effort has gone into proving that convergence to equilibrium takes place exponentially fast in time, under minimal conditions on the initial datum. See \cite{cl, cgt, dophd, grad, mouh} and the references therein. However, the mathematics needed to prove these last statements is generally quite complex. On the other hand, it is desirable to have also simple proofs of the asymptotic behavior even under the restrictive conditions mentioned at the beginning. From an historical point of view, the problem has been pointed out on page 345 of \cite{mck6}. Its usefulness has recently emerged, for example, in the proof of Theorem 1.2 in \cite{drPTRF}.

\subsection{The equation}

The equation under study in the present work is the \emph{homogeneous Boltzmann equation for Maxwellian molecules}, which is concerned with a spatially homogeneous dilute gas composed of a very large number of like particles.  See \cite{cerS, trumu, vil} for an exhaustive and detailed treatment of the well-known Boltzmann model. The locution ``Maxwellian molecules'' means that each collision is influenced by a repulsive force proportional to $r^{-5}$, $r$ being the distance between two colliding particles, a very particular situation which was studied for the first time in \cite{max}. In case of absence of external forces, the equation reads
\begin{equation} \label{eq:boltzmann}
\frac{\partial}{\partial t} f(\mathbf{v}, t) = Q_b[f(\cdot, t), f(\cdot, t)](\vb)
\end{equation}
with $(\vb, t)$ in $\rthree \times (0, +\infty)$. A solution of (\ref{eq:boltzmann}), $f(\cdot, t)$, is required to be a probability density function (pdf) in the first variable, at each instant $t$. The \emph{collision operator} $Q_b$ is defined for every pair $(\varphi, \psi)$ of real-valued functions in $\Lone$ through the relation
\begin{equation} \label{eq:Q}
Q_b[\varphi, \psi](\vb) := \int_{\mathbb{R}^3}\int_{S^2} [\varphi(\vb_{\ast}) \psi(\wb_{\ast}) \ - \ \varphi(\vb) \psi(\wb)] b\left(\frac{\wb - \vb}{|\wb - \vb|} \cdot \omb \right) \unifSo \ud \wb
\end{equation}
where $u_{S^2}$ stands for the uniform probability measure on the unit sphere $S^2$, embedded in $\rthree$. Moreover,
the \emph{post-collisional velocities} $\vb_{\ast}$ and $\wb_{\ast}$ must obey the conservation of momentum and kinetic energy, that is
$$
\vb + \wb = \ \vb_{\ast} + \wb_{\ast} \ \ \ \ \ \ \ \ \text{and} \ \ \ \ \ \ \ \
|\vb|^2 + |\wb|^2 = \ |\vb_{\ast}|^2 + |\wb_{\ast}|^2
$$
and, consequently, can be parametrized by unit vectors $\omb$ in $S^2$ according to
\begin{equation} \label{eq:velpost}
\begin{array}{lll}
\vb_{\ast} &= \ \vb &+ \ \ [(\wb - \vb) \cdot \omb] \ \omb  \\
\wb_{\ast} &= \ \wb &- \ \ [(\wb - \vb) \cdot \omb] \ \omb
\end{array}
\end{equation}
where $\cdot$ denotes the standard scalar product. The $[0, +\infty)$-valued function $b$ -- the so-called \emph{Maxwellian collision kernel} -- is defined on $(-1, 1)$ and carries the information about any single collision at a microscopic level. This function satisfies, for all $x$ in $(-1, 1)$, the symmetry condition
\begin{equation} \label{eq:bsymm}
b(x) = b(\sqrt{1 - x^2}) \frac{|x|}{\sqrt{1 - x^2}} = b(-x)
\end{equation}
and the so-called \emph{Grad angular cutoff}, here written as
\begin{equation} \label{eq:cutoff}
\int_{0}^{1} b(x) \ud x = 1 \ .
\end{equation}
See Section 3 in Chapter 2A of \cite{vil} for further information about collision kernels.

Existence and uniqueness for solutions of (\ref{eq:boltzmann}) are well-understood questions, at least when (\ref{eq:bsymm})-(\ref{eq:cutoff}) hold. More precisely, in \cite{mor} it is proved that, given a pdf $f_0$ as initial datum, the resulting Cauchy problem admits a unique solution $f(\cdot, t)$. Another question of some relevance for its mathematical and physical implications is that there exist non trivial stationary solutions of (\ref{eq:boltzmann}), that can be seen as the possible equilibrium distributions. Within the class of all pdf's on $\rthree$, these stationary solutions are exactly
the \emph{Maxwellian pdf's}, i.e.
$$
M_{\vb_0, \sigma}(\vb) := \left(\frac{1}{2\pi \sigma^2}\right)^{3/2} \exp\{-\frac{1}{2 \sigma^2} |\vb - \vb_0|^2\}
$$
where $(\vb_0, \sigma)$ varies in $\rthree \times (0, +\infty)$. See comments about Theorem 1.2 in \cite{drPTRF}. Relevant properties, from a physical point of view, are collected in Chapter VIII of \cite{trumu}.

\subsection{Approach to equilibrium based on the linearization of the collision operator}

The most important problem connected with the long-time behavior of the solutions of (\ref{eq:boltzmann}) is the quantification of the rate of convergence. From a historical point of view, the first technique introduced to pursue this goal was based on a linearization of the non-linear equation (\ref{eq:boltzmann}) and on the spectral analysis of the resulting linearized collision operator. The pioneering study explaining this line of reasoning is \cite{hil}, but met with little success. Indeed, it required too restrictive conditions on $f_0$ in order to obtain a quantification of the desired convergence which, in any case, was plainly improvable. This is the main reason for the introduction of alternative techniques. A remarkable exception is the recent work \cite{mouh}, in which the linearization technique has been reappraised and improved to obtain a definitive result on the rapidity of convergence to equilibrium of the solution of the spatially homogeneous Boltzmann equation with \emph{hard potentials}, which, however, does not include the present Maxwellian case.

It is worth providing here a self-contained treatment of the linearization procedure, a subject which is still scattered, sometimes with discordant notation, in different sources. See, for example, \cite{grad, grulin}.

It will be assumed that the initial datum $f_0$ satisfies
\begin{equation} \label{eq:normalizations}
\intethree \vb f_0(\vb) \ud\vb = \mathbf{0} \ \ \ \ \text{and} \ \ \ \ \intethree |\vb|^2 f_0(\vb) \ud\vb = 3 \ .
\end{equation}
From conservation of momentum and kinetic energy, namely
\begin{eqnarray}
\intethree \vb f(\vb, t) \ud\vb &=& \intethree \vb f_0(\vb) \ud\vb \nonumber \\
\intethree |\vb|^2 f(\vb, t) \ud\vb &=& \intethree |\vb|^2 f_0(\vb) \ud\vb \nonumber
\end{eqnarray}
for every $t$ in $[0, +\infty)$, it follows that (\ref{eq:normalizations}) can be assumed without any loss of generality. Moreover, under (\ref{eq:normalizations}), the above conservations are preserved in the limit and the relative Maxwellian equilibrium turns out to be $M_{\mathbf{0}, 1}$, which will be simply indicated with $M$.

A central role will be played throughout this work by the so-called \emph{linearized collision operator} $L_b$ defined by
\begin{eqnarray}
L_b[h](\vb) &:=& \intethree\int_{S^2} M(\wb) \ [h({\vb_{\ast}}) +
h({\wb_{\ast}}) \ - \ h(\vb) - h(\wb)] \times \nonumber \\
&\times& b\left(\frac{\wb - \vb}{|\wb - \vb|} \cdot \omb \right) \unifSo \ud \wb \ . \nonumber
\end{eqnarray}
The introduction of $L_b$ can be justified as follows: After the substitution $f(\vb, t) = M(\vb) (1 + h(\vb, t))$, equation (\ref{eq:boltzmann}) changes into a new equation for $h$, namely
\begin{equation} \label{eq:boltzmannh}
\frac{\partial}{\partial t} h(\vb, t) = L_b[h(\cdot, t)](\vb) + R_b[h(\cdot, t), h(\cdot, t)](\vb)
\end{equation}
with $R_b$ defined by
\begin{equation} \label{eq:21cls}
R_b[\varphi, \psi](\vb) := \frac{1}{M(\vb)} Q_b[M \varphi, M \psi](\vb) \ .
\end{equation}
At this stage, the function $h$ can be thought of as a sort of ``remainder'', which becomes smaller and smaller when $t$ increases. Therefore, if the contribution of the quadratic operator $R_b$ in (\ref{eq:boltzmannh}) becomes negligible with respect to that given by $L_b$, then the spectral properties of $L_b$ could provide a quantitative information about the rapidity of convergence of $h$ to the null function. This insight, which may lead to the desired conclusion of quantifying the convergence to equilibrium, actually sums up the very content of this note and will be formalized in a rigorous way in the sequel. Moreover, it is crucial to point out that this strategy works only under some restrictions on $f_0$, to be specified as well. As to the mentioned spectral analysis of $L_b$, it can be a very difficult task if based on the natural domain of $L_b$, namely the space of functions $h : \rthree \rightarrow \rone$ which can be written as $\frac{f(\cdot)}{M(\cdot)} - 1$, when $f$ is any pdf on $\rthree$. A remarkable idea in \cite{hil} consists in the introduction of the Hilbert space $\mathcal{H} := \mathrm{L}^2(\rthree, M(\xb)\ud \xb)$ as a new domain for $L_b$, a device to makes computations feasible, since there is a Fourier basis for $\mathcal{H}$ that diagonalizes $L_b$. To complete the necessary notation, introduce $(\cdot, \cdot)_{\ast}$ and $\lnorm \cdot \rnorma$ to denote the scalar product and the norm of $\mathcal{H}$, respectively, and $\mathcal{N}_{\delta} := \{h \in \mathcal{H} \ | \ \lnorm h \rnorma \ \leq \delta\}$ to indicate the ball of radius $\delta$ centered at the origin. The kernel of $L_b$ coincides with the five-dimensional linear subspace $\text{span}\{1, v_1, v_2, v_3, |\vb|^2\}$ generated by the \emph{collisional invariants}. $\mathcal{H}_0$ will indicate the orthogonal complement in $\mathcal{H}$ of the kernel of $L_b$. Since
\begin{eqnarray}
\intethree \vb f(\vb, t) \ud\vb &=& \intethree \vb M(\vb) \ud\vb \nonumber \\
\intethree |\vb|^2 f(\vb, t) \ud\vb &=& \intethree |\vb|^2 M(\vb) \ud\vb \nonumber
\end{eqnarray}
for every $t$ in $[0, +\infty)$, it follows that if $h(\cdot, t)$ belongs to $\mathcal{H}$ then it is in the subspace $\mathcal{H}_0$ for all $t$. On the new domain, the linear operator $L_b$ is self-adjoint and negative with a discrete set of eigenvalues, the least negative of which, $\Lambda_b$, represents the \emph{spectral gap}. A precise analysis is contained in \cite{do}, where it is also shown that
$$
\Lambda_b = -2 \int_{0}^{1} x^2(1 - x^2) b(x) \ud x
$$
and, for every $\varphi$ in $\mathcal{H}_0$,
\begin{equation} \label{eq:spectralinequality}
\big{(}L_b[\varphi], \varphi\big{)}_{\ast} \leq \Lambda_b \lnorm \varphi \rnorma^2 \ .
\end{equation}
This spectral gap has been considered as a reference value for the rate of exponential convergence of $f(\cdot, t)$ to $M$ in the original equation, but this claim held out long as an unproved conjecture. The difficult point consists in the fact that the spectral properties of $L_b$, viewed as an operator on $\mathcal{H}$, are not directly connected with the properties of the solution of the non-linear equation (\ref{eq:boltzmann}), where it would be more natural to consider the $\mathrm{L}^1$ distance. In point of fact, a definitive result has been recently obtained in \cite{dophd}, but the proof is quite long and complex. With a view to simplified, but rigorous, treatments of the subject one is led to renounce the pursuit of optimal rates in exchange for significant simplifications of the mathematical proof. The following result just goes in this direction. \begin{thm} \label{thm:opt}
\emph{Let} (\ref{eq:bsymm})-(\ref{eq:cutoff}) \emph{be in force and let} $\delta := |\Lambda_b|/16$. \emph{If}
\begin{equation} \label{eq:zero}
\frac{f_0(\cdot) - M(\cdot)}{M(\cdot)} \in \mathcal{N}_{\delta} \ ,
\end{equation}
\emph{then}
\begin{equation} \label{eq:initialcls}
\frac{f(\cdot, t) - M(\cdot)}{M(\cdot)} \in \mathcal{N}_{2\delta}
\end{equation}
\emph{holds true for all times} $t > 0$. \emph{Moreover, under} (\ref{eq:zero}),
\begin{equation} \label{eq:cls}
\lnorm f(\cdot, t) - M(\cdot) \rnormm := \intethree \big{|} f(\vb, t) - M(\vb) \big{|} \ud\vb\ \ \leq\ C_{\ast} e^{\frac{1}{2}\Lambda_b t}
\end{equation}
\emph{is valid with}
$$
C_{\ast} := \left(\frac{1}{\lnorm h(\cdot, 0) \rnorm_{\ast}} + \frac{2}{\Lambda_b} \right)^{-2} \ .
$$
\end{thm}
\vspace{5mm}
With reference to the main motivation for the present note explained at the beginning, it is worth indicating how Theorem \ref{thm:opt} is applied in the proof of Theorem 2.1 of \cite{drPTRF}. The problem that one must tackle therein reduces to the case in which $f_0(\vb) = \prod_{i = 1}^{3} g_{\sigma_i}(v_i)$, where $g_{\sigma}(x) := \frac{1}{\sqrt{2\pi\sigma^2}} \exp\{-\frac{1}{2\sigma^2}x^2\}$ and $\sum_{i = 1}^{3} \sigma_{i}^{2} = 3$. Indeed,
\begin{eqnarray}
\Big{|}\Big{|} \frac{f_0(\cdot) - M(\cdot)}{M(\cdot)} \Big{|}\Big{|}_{\ast}^2 &\leq& 3 \frac{1}{\sigma_2 \sqrt{2 - \sigma_{2}^{2}}} \frac{1}{\sigma_3 \sqrt{2 - \sigma_{3}^{2}}} \left[\frac{1}{\sigma_1 \sqrt{2 - \sigma_{1}^{2}}} - 1\right] \nonumber \\
&+& 3 \frac{1}{\sigma_3 \sqrt{2 - \sigma_{3}^{2}}} \left[\frac{1}{\sigma_2 \sqrt{2 - \sigma_{2}^{2}}} - 1\right] \nonumber \\
&+& 3 \left[\frac{1}{\sigma_3 \sqrt{2 - \sigma_{3}^{2}}} - 1\right] \nonumber
\end{eqnarray}
whence condition (\ref{eq:zero}) plainly follows provided that
$$
\sigma_{i}^{2} \in \left[1 - \frac{\sqrt{42 + \delta^2}}{21 + \delta^2} \delta, 1 + \frac{\sqrt{42 + \delta^2}}{21 + \delta^2} \delta \right]
$$
for $i = 1, 2, 3$. \\

\section{Proof of the theorem}

The proof is split into two subsections: The former contains a discussion about the validity of condition (\ref{eq:initialcls}), the latter includes the proof of (\ref{eq:cls}).

\subsection{Existence and uniqueness near the equilibrium} \label{sect:exun}

Here the validity of (\ref{eq:initialcls}) is derived the study of equation (\ref{eq:boltzmannh}). Existence and uniqueness are tackled according to an approach rather different from the classical one presented in \cite{mor}, which requires new proofs. Following \cite{cls}, after fixing the initial datum $h_0$ in $\mathcal{H}_0$, the solution of the Cauchy problem,
resulting from (\ref{eq:boltzmannh}) and this initial condition, is meant as an element of $\mathrm{C}([0, \infty); \mathcal{H}_0) \cap \mathrm{C}^1([0, \infty); \mathcal{H}_0)$.

To start, let $\mathscr{T}^{t}$ denote the semigroup of linear operators on $\mathcal{H}_0$ sending an element $g$ onto the solution $\mathscr{T}^{t}[g]$ of the evolution equation $\frac{\partial}{\partial t} h(\vb, t) = L_b[h(\cdot, t)](\vb)$. It is well-known that $\mathscr{T}^{t}$ admits a characterization in the form of exponential semigroup $\exp\{t L_b\}$. The basic properties of $L_b$, collected, for example, in \cite{do}, guarantee that $\mathscr{T}^{t}[g]$ is actually an element of $\mathcal{H}_0$ whenever $g$ is in the same space. Following general references on abstract differential equations like \cite{lak, mar}, one obtains that the solution of (\ref{eq:boltzmannh}) admits the representation
\begin{equation} \label{eq:volterra}
h(\vb, t) = \mathscr{T}^{t}[h_0](\vb) + \int_{0}^{t} \mathscr{T}^{t-s}[R_b[h(\cdot, s), h(\cdot, s)]](\vb) \ud s
\end{equation}
which lends itself to be interpreted as a fixed point problem. \\

Now, a first preliminary fact, which follows from (\ref{eq:spectralinequality}), is that
\begin{equation} \label{eq:contraction}
\lnorm \mathscr{T}^{t}[g] \rnorma\ \leq e^{\Lambda_b t} \lnorm g \rnorma
\end{equation}
for every $g$ in $\mathcal{H}_0$ and all $t$ in $[0, +\infty)$. Another preliminary fact is encompassed in the inequality
\begin{equation} \label{eq:29cls}
|(R_b[\varphi, \psi], \rho)_{\ast}|\ \leq\ 2 \lnorm \varphi \rnorma  \lnorm \psi \rnorma  \lnorm \rho \rnorma
\end{equation}
which is valid for every $\varphi, \psi$ and $\rho$ in $\mathcal{H}$. A direct consequence of (\ref{eq:29cls}) is
\begin{equation} \label{eq:24cls}
\lnorm R_b[\varphi, \psi]\rnorma\ \leq 2 \lnorm \varphi \rnorma  \lnorm \psi \rnorma \ .
\end{equation}
To prove (\ref{eq:29cls}), it can be observed that the quantity
\begin{eqnarray}
&{}& (R_b[\varphi, \psi], \rho)_{\ast} = \intethree \rho(\vb) Q_b[M \varphi, M \psi](\vb) \ud \vb \nonumber \\
&=& \intethree \intethree \int_{S^2} \rho(\vb) \varphi(\vb_{\ast}) \psi(\wb_{\ast}) M(\vb_{\ast}) M(\wb_{\ast}) b\left(\frac{\wb - \vb}{|\wb - \vb|} \cdot \omb \right) \unifSo \ud \vb \ud \wb \nonumber \\
&-& \intethree \intethree \int_{S^2} \rho(\vb) \varphi(\vb) \psi(\wb) M(\vb_{\ast}) M(\wb_{\ast}) b\left(\frac{\wb - \vb}{|\wb - \vb|} \cdot \omb \right) \unifSo \ud \vb \ud \wb \nonumber \\
\label{eq:split1}
\end{eqnarray}
is decomposed as difference of two terms. The former, which reads
\begin{eqnarray}
&{}& \intethree \intethree \int_{S^2} \left[\rho(\vb) M^{1/2}(\vb_{\ast}) M^{1/2}(\wb_{\ast}) b^{1/2}\left(\frac{\wb - \vb}{|\wb - \vb|} \cdot \omb \right)\right] \times \nonumber \\
&\times& \left[\varphi(\vb_{\ast}) \psi(\wb_{\ast}) M^{1/2}(\vb_{\ast}) M^{1/2}(\wb_{\ast}) b^{1/2}\left(\frac{\wb - \vb}{|\wb - \vb|} \cdot \omb \right)\right] \unifSo \ud \vb \ud \wb \ , \nonumber
\end{eqnarray}
can be bounded from above, by means of the Cauchy-Schwarz inequality, by
\begin{eqnarray}
&{}& \left[ \intethree \intethree \int_{S^2}\rho^2(\vb) M(\vb_{\ast}) M(\wb_{\ast}) b\left(\frac{\wb - \vb}{|\wb - \vb|} \cdot \omb \right)\unifSo \ud \vb \ud \wb \right]^{1/2} \times \nonumber \\
&\times& \left[\intethree \intethree \int_{S^2} \varphi^2(\vb_{\ast}) \psi^2(\wb_{\ast}) M(\vb_{\ast}) M(\wb_{\ast}) b\left(\frac{\wb - \vb}{|\wb - \vb|} \cdot \omb \right) \unifSo \ud \vb \ud \wb \right]^{1/2}\ . \nonumber
\end{eqnarray}
Since $M(\vb_{\ast}) M(\wb_{\ast}) = M(\vb) M(\wb)$ and $\int_{S^2} b(\ub \cdot \omb) \unifSo = 1$ for every $\ub$ in $S^2$,
it follows that
$$
\left[ \intethree \intethree \int_{S^2}\rho^2(\vb) M(\vb_{\ast}) M(\wb_{\ast}) b\left(\frac{\wb - \vb}{|\wb - \vb|} \cdot \omb \right)\unifSo \ud \vb \ud \wb \right]^{1/2} =\ \lnorm \rho \rnorma \ .
$$
Moreover, since $(\vb, \wb) \mapsto (\vb_{\ast}, \wb_{\ast})$ is a linear isometry of $\rone^6$ for every $\omb$ in $S^2$ and
$$
b\left(\frac{\wb - \vb}{|\wb - \vb|} \cdot \omb \right) = b\left(\frac{\wb_{\ast} - \vb_{\ast}}{|\wb_{\ast} - \vb_{\ast}|} \cdot \omb \right) \ ,
$$
the change-of-variable theorem yields
\begin{eqnarray}
&{}& \left[\intethree \intethree \int_{S^2} \varphi^2(\vb_{\ast}) \psi^2(\wb_{\ast}) M(\vb_{\ast}) M(\wb_{\ast}) b\left(\frac{\wb - \vb}{|\wb - \vb|} \cdot \omb \right) \unifSo \ud \vb \ud \wb \right]^{1/2} \nonumber \\
&=& \lnorm \varphi \rnorma  \lnorm \psi \rnorma \ , \nonumber
\end{eqnarray}
which is the desired bound for the former term under discussion. Then, since $\int_{S^2} b(\ub \cdot \omb) \unifSo = 1$ for every $\ub$ in $S^2$, the latter term in (\ref{eq:split1}) is equal to
$$
\intethree M(\vb) \varphi(\vb) \rho(\vb) \ud \vb \cdot \intethree M(\vb) \psi(\vb) \ud \vb = (\varphi, \rho)_{\ast} \cdot (\psi, 1)_{\ast}
$$
and the Cauchy-Schwarz inequality gives $|(\varphi, \rho)_{\ast} \cdot (\psi, 1)_{\ast}| \leq\ \lnorm \varphi \rnorma  \lnorm \psi \rnorma  \lnorm \rho \rnorma$. The proof of (\ref{eq:29cls}) follows from the combination of the upper bounds just obtained. \\

After these preliminaries, existence and uniqueness will be proved via a contraction mapping principle, as in \cite{cls}.
The first step consists in the definition, for $T > 0$, of the Banach space $\mathbb{X} := \mathrm{C}([0, T]; \mathcal{H}_0)$ endowed with the norm
$$
\tlnorm x \trnorm\ := \sup_{t \in [0, T]} \lnorm x(t) \rnorma \ .
$$
Then, the formula
$$
Z_b[x] := \mathscr{T}^{t}[h_0] + \int_{0}^{t} \mathscr{T}^{t-s}[R_b[x(s), x(s)]] \ud s
$$
defines an operator on $\mathbb{X}$. Indeed both $h_0$ and $x(s)$ belong to $\mathcal{H}_0$ and, consequently, $R_b[x(s), x(s)]$ and $\mathscr{T}^u[R_b[x(s), x(s)]]$ are again elements of $\mathcal{H}_0$, for every $s, u$ in $[0, T]$. After setting $D := \{x \in \mathbb{X} \ | \ \tlnorm x \trnorm\ \leq |\Lambda_b|/8\}$, which is obviously a closed subset of $\mathbb{X}$, it can be proved that $Z_b(D) \subset D$ and that
\begin{equation} \label{eq:contrazione}
\tlnorm Z_b[x] - Z_b[y] \trnorm\ \leq \frac{1}{2}\tlnorm x - y \trnorm
\end{equation}
for every $x$ and $y$ in $D$, provided that $h_0$ belongs to $\mathcal{N}_{\delta}$ with $\delta = |\Lambda_b|/16$.

The proof of the first claim is based on (\ref{eq:contraction}) and (\ref{eq:24cls}), which give
\begin{eqnarray}
\tlnorm Z_b[x] \trnorm\ &\leq& \sup_{t \in [0, T]} \left[\lnorm h_0 \rnorma e^{\Lambda_b t} + \int_{0}^{t} e^{\Lambda_b (t-s)} \lnorm R_b[x(s), x(s)] \rnorma \ud s\right] \nonumber \\
&\leq& \lnorm h_0 \rnorma\ + 2  \sup_{t \in [0, T]} \int_{0}^{t} e^{\Lambda_b (t-s)} \lnorm x(s)\rnorma^2 \ud s \ . \nonumber
\end{eqnarray}
In view of the bound on $h_0$ and the fact that $x$ is in $D$, the claim is proved by means of the inequality
$$
\tlnorm Z_b[x] \trnorm\ \leq\ \frac{1}{16}|\Lambda_b|\ +\ 2 \frac{1}{64}|\Lambda_b|^2 \frac{1}{|\Lambda_b|}\ <\ \frac{1}{8}|\Lambda_b| \ .
$$
To verify (\ref{eq:contrazione}), inequalities (\ref{eq:contraction}) and (\ref{eq:24cls}) can be used again to write
\begin{eqnarray}
\tlnorm Z_b[x] - Z_b[y]\trnorm\ &\leq& \sup_{t \in [0, T]} \int_{0}^{t} e^{\Lambda_b (t-s)} \lnorm R_b[x(s), x(s)] -  R_b[y(s), y(s)] \rnorma \ud s \nonumber \\
&=& \sup_{t \in [0, T]} \int_{0}^{t} e^{\Lambda_b (t-s)} \lnorm R_b[x(s) + y(s), x(s) - y(s)] \rnorma \ud s \nonumber \\
&\leq& \sup_{t \in [0, T]} 2\int_{0}^{t} e^{\Lambda_b (t-s)} \lnorm x(s) + y(s) \rnorma \cdot \lnorm x(s) - y(s) \rnorma \ud s \nonumber \\
&\leq& 2 \frac{1}{4}|\Lambda_b| \frac{1}{|\Lambda_b|} \tlnorm x - y \trnorm \ . \nonumber
\end{eqnarray}
Existence and uniqueness of the solution of (\ref{eq:boltzmannh}) are now direct consequences of Theorem 1.1 in Chapter IV of \cite{mar}, and this solution can be viewed as a map from $[0, T]$ into $\mathcal{H}_0$. Finally, since the above argument is independent of the choice of $T$, the solution can be extended to $[0, +\infty)$ and this proves the first part of Theorem \ref{thm:opt}. Indeed, the validity of (\ref{eq:initialcls}) is nothing but the translation of the fact that the fixed point problem represented by (\ref{eq:volterra}) admits a unique solution in $D$, rewritten, through the equation $f(\cdot, t) = M(\cdot) \big{(}1 + h(\cdot, t)\big{)}$, in terms of $f(\cdot, t)$.

\subsection{Rapidity of convergence to equilibrium} \label{sect:convergence}

This section contains the proof of (\ref{eq:cls}). Starting from $f(\vb, t) = M(\vb) (1 + h(\vb, t))$, the Jensen inequality entails
\begin{equation} \label{eq:CK}
\lnorm  f(\vb, t) - M(\vb) \rnormm^2 \ \leq \ \intethree \frac{\left(f(\vb, t) - M(\vb)\right)^2}{M(\vb)} \ud\vb =\ \lnorm h(\cdot, t) \rnorma^2 \ =: \theta(t)\ .
\end{equation}
Taking the scalar product $(\cdot, \cdot)_{\ast}$ of both members of (\ref{eq:boltzmannh}) with the solution $h(\cdot, t)$ of the same equation yields
$$
\frac{\ud}{\ud t} \big{(} h(\vb, t), h(\vb, t)\big{)}_{\ast} = \big{(}L_b[h(\cdot, t)], h(\vb, t)\big{)}_{\ast} + \big{(}R_b[h(\cdot, t), h(\cdot, t)], h(\vb, t)\big{)}_{\ast} \ .
$$
Since $h(\cdot, t)$ belongs to $\mathcal{H}_0$ for every $t \geq 0$, (\ref{eq:spectralinequality}) and (\ref{eq:29cls}) lead to
$$
\frac{\ud}{\ud t} \theta(t) \leq \Lambda_b \theta(t) + 2 [\theta(t)]^{3/2} \ .
$$
After setting $\vartheta(t) := \theta(t) e^{-\Lambda_b t}$, the above inequality becomes
$$
\frac{\ud}{\ud t} \vartheta(t) \leq 2 [\vartheta(t)]^{3/2} e^{-\Lambda_b t} = 2 [\vartheta(t)]^{3/2} e^{\frac{1}{2} \Lambda_b t} \ .
$$
Whence,
$$
-2\left[\frac{1}{\sqrt{\vartheta(t)}} - \frac{1}{\sqrt{\vartheta(0)}}\right] = \int_{0}^{t} [\vartheta(\tau)]^{-3/2} \vartheta^{'}(\tau) \ud \tau
\leq 2 \int_{0}^{t} e^{\frac{1}{2} \Lambda_b \tau} \ud \tau \leq  \left(\frac{4}{-\Lambda_b}\right) \nonumber
$$
and, after some elementary algebra,
\begin{eqnarray}
\vartheta(t) &\leq& \left[\frac{1}{\sqrt{\vartheta(0)}} + \frac{2}{\Lambda_b}\right]^{-2} = \left[\frac{1}{\sqrt{\theta(0)}} + \frac{2}{\Lambda_b}\right]^{-2} \nonumber \\
&=& \left[\frac{1}{\lnorm h(\cdot, 0) \rnorma} + \frac{2}{\Lambda_b}\right]^{-2} = C_{\ast} \ . \label{eq:PSI}
\end{eqnarray}
Note that (\ref{eq:initialcls}) guarantees that $C_{\ast}$ is a well-defined, strictly positive real constant. Combination of (\ref{eq:PSI}) with the definition of $\vartheta$ gives $\theta(t) \leq C_{\ast} e^{\Lambda_b t}$, which is the desired conclusion.

\vspace{1cm}

\footnotesize{\textsc{emanuele dolera \\
dipartimento di matematica pura e applicata ``giuseppe vitali'' \\
universit\`a degli studi di modena e reggio emilia \\
via campi 213/b, 41100 modena, italy \\
e-mail:} emanuele.dolera@unimore.it, emanuele.dolera@unipv.it}

\end{document}